\magnification=\magstep1
\tolerance=500
\rightline{20 April, 2020}
\bigskip
\centerline{\bf Fourier Transform, Quantum Mechanics and Quantum Field Theory}
\centerline{\bf on the Manifold of General Relativity}
\bigskip
\centerline{ L.P. Horwitz}
\bigskip
\centerline{ School of Physics, Tel Aviv University, Ramat Aviv
69978, Israel}
\centerline{ Department of Physics, Bar Ilan University, Ramat Gan
52900, Israel}
\centerline{ Department of Physics, Ariel University, Ariel 40700, Israel}
\bigskip

\noindent email: larry@tauex.tau.ac.il
\bigskip
\noindent{\it Abstract}
\smallskip
\par A proof is given for the Fourier transform for functions in a quantum mechanical Hilbert space on a non-compact manifold in general relativity. In the (configuration space) Newton-Wigner representation we discuss the spectral decomposition of the canonical operators and give a proof of the Parseval-Plancherel relation and the Born rule for linear superposiion. We then discuss the representations of pure quantum states and their dual vectors, and construct the Fock space and the associated quantum field theory for Bose-Einstein and Fermi-Dirac statistics.

\bigskip
\noindent Keywords: Fourier transform, manifold, general relativity, Hilbert space, canonical operators, quantum theory, quantum field theory on curved spacetime
\bigskip

\noindent PACS: 04.20.Cv, 04.20.Fy, 04.20.-q, 04.20.Ds, 04.62.+v

\bigskip
\noindent{\bf 1. Fourier analysis}
\bigskip
\par In a recent paper[1] discussing the embedding of the relativistic canonical classical and quantum theory of Stueckelberg, Horwitz and Piron[2][3][4] (see also [5][6]) into general relativity, the Fourier transform on the manifold, necessary for the construction of a canonical quantum theory, was introduced without proof. In this paper we provide a proof and clarify the conditions for its validity\footnote{*}{Note that Abraham, Marsden and  Ratiu [7] apply the formal Fourier transform on a manifold in three dimensions without proof.}. We then discuss the configuration space Newton-Wigner representation [1][8] and the spectral decomposition of the canonical operators. The Dirac form of the quantum theory leads to the notion of a dual space (which does not coincide with the complex conjugate). 
We give a proof of the Parseval-Plancheret theorem[9][10] and discuss the Born rule for linear superpositions. With the help of the definition of the dual states, we then construct the Fock space and formulate the associated quantum field theory. 
\par Since we are working in a canonical framework[1] we do not make use of eigenfunctions of the Laplace-Beltrami operator [11]. Although there is a translation group along geodesic curves generated by the canonical momentum, we shall not be concerned here with the general problem of Fourier analysis on group manifolds.  
\par We wish to study the construction of the Fourier transform on a manifold with metric $g_{\mu\nu}(x)$ ( $x \equiv x^\mu,\ \ \mu=(0,1,2,3)$, local flat space metric $(-,+,+,+)$) and (compatible) connection form ${\Gamma^\lambda}_{\mu\nu}(x)$. We shall assume that the manifold is non-compact and geodesically complete.  
\par For a function $f(x)$  defined almost everywhere on the manifold $\{x\}$, we define the Fourier transform[1]
$$ {\tilde f}(p) = \int d^4x \sqrt{g}\  e^{-ip_\mu x^\mu} f(x), \eqno(1)$$
where $g = -\det g_{\mu\nu}$ and the integral is carried out (in the Riemannian sense) in the limit of the sum over small spacetime volumes with invariant measure
$d^4x \sqrt{g}$.  Note that $p_\mu x^\mu\equiv -p_0 x^0 + p_1 x^1 +p_2 x^2 +p_3 x^3$ is not local diffeomorphism invariant, and hence not a scalar product, on the manifold. The Fourier transform as we have defined it is carried out in the framework of a given, arbitrary, coordinatization.  
\par Provided that
$$ \int d^4p \ e^{-ip_\mu(x^\mu-x'^\mu)} = (2\pi)^4 \delta^4(x-x'), \eqno(2)$$
so that
$$(2\pi)^{-4}\int d^4 x'\int d^4p\  e^{-ip_\mu(x^\mu-x'^\mu)} = 1, \eqno(3)$$
to prove consistency of the definition $(1)$ and the requirement $(2)$, we must have \footnote{**}{It is shown in [1] that, in particular,  the statement $(4)$ is covariant under local diffeomorphisms, {\it i.e.}, under a local change of coordinatization.}   $$ {\tilde f}(p)= {1 \over (2\pi)^4} \int d^4x \int d^4p'\  e^{-i(p_\mu -p'_\mu)x^\mu} {\tilde f}(p').
  \eqno(4)$$
  This condition follows by taking the inverse transform of $(1)$, {\it i.e.},
  $$\int d^4 p \ e^{ip_\mu x^\mu} {\tilde f}(p) = (2\pi)^4 \sqrt{g(x)} f(x),\eqno(5)$$
  so that
  $$f(x) = {1 \over (2\pi)^4} {1\over \sqrt{g(x)}}\int d^4 p\  e^{ip_\mu {x^\mu}} {\tilde f}(p).\eqno(6)$$
  Substituting this result into $(1)$, one obtains the condition $(4)$.
 \par   Exchanging the order of integrations in $(4)$,assuming convergence in  either order, we see that we must study the function (in a particular coordinatization $\{x\}$ and cotangent space $\{p\}$)
       $$\Delta(p-p') \equiv {1 \over (2\pi)^4} \int d^4x\  e^{-i(p_\mu -p'_\mu) x^\mu}  \eqno(7)$$
which should act as the distribution $ \delta^4(p-p')$.
\par To prove this consistency condition, following the method of Reed and Simon[12] in their discussion of Lebesgue integration, we represent the integral as a sum over small boxes around the set of points $\{x_B\}$ that cover the space (which we have assumed to be non-compact), and eventually take the limit as for a Riemann-Lebesgue integral. In each small box, the coordinatization arises from an invertible transformation from the local tangent space in that neighborhood. We write
$$ x^\mu = {x_B}^\mu + \eta^\mu \ \ \ \in {\rm box B} \eqno(8)$$
where
$$ \eta^\mu =  {\partial x^\mu \over \partial \xi^\lambda} \xi^\lambda  \eqno(9)$$
and $\xi^\lambda$ (small) is in the flat local tangent space at $x_B$.
\par We now write the integral $(7)$ as
$$\eqalign{\Delta(p-p') &= {1 \over (2\pi)^4} \Sigma_B \int_B d^4 \eta\  e^{-i(p_\mu -p'_\mu)({x_B}^\mu +\eta^\mu)}\cr
&= {1 \over (2\pi)^4} \Sigma_B e^{-i(p_\mu -p'_\mu){x_B}^\mu} \int_B d^4 \eta \ e^{-i(p_\mu -p'_\mu)\eta^\mu}. \cr} \eqno(10)$$
Let us call the measure at $B$ 
$${\bf\Delta} \mu(B, p-p')= \int_B d^4 \eta \ e^{-i(p_\mu -p'_\mu)\eta^\mu} . \eqno(11)$$
 In this neighborhood, define
$$ {\partial x^\mu \over \partial \xi^\lambda} = {\partial \eta^\mu \over \partial \xi^\lambda}\equiv {a^\mu}_\lambda (B), \eqno(12)$$
which may be taken to be a constant matrix in each small box. In $(11)$, we then have
$${\bf\Delta} \mu(B, p-p')= \det a \int_B d^4 \xi \  e^{-i(p_\mu -p'_\mu){a^\mu}_\lambda (B)\xi^\lambda}. \eqno(13)$$
\par We now make a change of variables for which ${\xi'}^\mu={a^\mu}_\lambda (B)\xi^\lambda$; then, since $d^4\xi' = {\det a}d^4\xi $, we have
$${\bf\Delta} \mu(B, p-p') = \int_{B'(B)} d^4 \xi'\  e^{i(p_\mu -p'_\mu)\xi^{\mu '}} . \eqno(14)$$
in each box.
 We remark that the local coordinate transformation in each box $B$ results, locally, in the metric $g_{\lambda\sigma}(B)=\eta_{\mu\nu} {\partial \xi^\mu\over \partial\eta^\lambda}{\partial \xi^\nu\over \partial\eta^\sigma}$ so that $\det g(B)= -(\det a)^{-2}$, and therefore
 $\det a$ is well-defined. 
\par  However, the transformation ${a^\mu}_
\lambda (B)$ in the neighborhood of each point $B$ is, in general, different, and therefore the set of transformed boxes may not cover (boundary deficits) the full domain of spacetime coordinates. It is easy to show, in fact, that the contribution of the boundary deficits of this naive partition of the spacetime may diverge. 
\par We may avoid this problem using our assumption of geodesic completeness of the manifold and taking the covering set of boxes along geodesic curves. Parallel transport of the tangent space boxes then fills the space in the neighborhood of the geodesic curve we are following, and each infinitesimal box carries an invariant volume (Liouville type flow) transported along a geodesic curve. For succesive boxes along the geodesic curve, since the boundaries are determined by parallel transport (rectilinear shift in the succession of local tangent spaces), there is no volume deficit between adjacent boxes.
\par We may furthermore translate a geodesic curve to an adjacent geodesic by the mechanism discussed in [13], so that boxes associated with adjacent geodesics are also related by parallel transport. In this way, we may fill the entire (geodesically complete) spacetime volume. 
\par We may then write $(10)$ as
$$\Delta(p-p')= {1 \over (2\pi)^4} \Sigma_B \ e^{-i(p_\mu -p'_\mu){x_B}^\mu}{\bf\Delta} \mu(B, p-p'), \eqno(15)$$
\par Our construction has so far been based on elements constructed in the tangent space in the neighborhood B of each point $x_B$. Relating all points along a geodesic to the corresponding local tangent spaces, and putting each patch in correspondence by continuity, we may consider the set $\{x_B\}$ to be in correspondence with an extended flat space $\{\xi(x_B)\}$,  to obtain\footnote{***}{This procedure is somewhat similar to the method followed in the simpler case of constant curvature by Georgiev[14] who, however, used eigenvalues of the Laplace-Beltrami operator.} 
$$\Delta(p-p')= {1 \over (2\pi)^4} \Sigma_B \ e^{-i(p_\mu -p'_\mu){\xi_B}^\mu}{\bf \Delta} \mu(\xi_B, p-p'). \eqno(16)$$
In the limit of small spacetime box volume, this approaches a Lebesgue type
integral on a flat space
$$\Delta(p-p')= {1 \over (2\pi)^4} \int e^{-i(p_\mu -p'_\mu){\xi}^\mu} d\mu(\xi, p-p'). \eqno(17)$$
If the measure is differentiable, we could write,
$$d\mu(\xi, p-p')= m(\xi, p-p') d^4\xi. \eqno(18)$$
\par In the small box, say, size $\epsilon$,
$$\eqalign{{\bf\Delta} \mu(B, p-p')  =\int_{-\epsilon/2}^{\epsilon/2} d\xi^0d\xi^1d\xi^2d\xi^3 \ e^{-i(p_\mu -p'_\mu)\xi^\mu} &= (2i)^4\Pi_{j=0}^{j=3}{\sin (p_j-p'_j){\epsilon\over 2} \over (p_j-p'_j)}\cr
&\rightarrow \epsilon^4\sim d^4\xi,\cr} \eqno(19)$$
so that for $\epsilon$ sufficiently small, $m(\xi, p-p')=1$, and we have
$$\Delta(p-p')= {1 \over (2\pi)^4} \int e^{-i(p_\mu -p'_\mu){\xi}^\mu} d^4\xi, \eqno(20)$$
or
$$\Delta(p-p')= \delta^4(p-p'). \eqno(21)$$
\par It is clear that the assertion $(19)$ requires some discussion. For $\epsilon \rightarrow 0$ we must be sure that $p'$ does not become too large. In each of the dimensions, what we want to find are conditions for which, in $(19)$,
$$ {sin p\epsilon\over p} \rightarrow \epsilon \eqno(22)$$
for $\epsilon \rightarrow 0$, where we have written $p$ for $p-p'$. Since the kernel $\Delta(p-p')$ is to act on elements of a Hilbert space $\{ {\tilde f}(p)\}$, the support for $p' \rightarrow \infty$ vanishes, so that $p-p'$ is essentially bounded.
\par As a distribution, on smooth functions $g(p)$, the left member of $(22)$ acts as
$$ G(\epsilon) \equiv \int_{-\infty}^\infty dp\  {sin p\epsilon\over p}g(p).\eqno(23)$$
The function $G(\epsilon)$ is analytic if $p^ng(p)$ has a Fourier transform for all $n$, since
$G(0)$ is identically zero, and successive derivatives correspond to the Fourier transforms of $p^ng(p)$ (differentiating under the integral). This implies, as a simple sufficient condition, that the (usual) Fourier transform of $g(p)$ is a $C^\infty$ function in the local tangent space $\{\xi\}$.  In this case we can reliably use the first order term in the Taylor expansion,
$$ {d \over d\epsilon} G(\epsilon)|_{\epsilon =0} = \int dp\  cos{\epsilon p}\ g(p)|_{\epsilon = 0} \eqno(24)$$
so that, for $\epsilon \rightarrow 0$,
$$ G(\epsilon) \rightarrow \epsilon {\tilde g}(0), \eqno(25)$$
where $ {\tilde g}(\xi)$ is the Fourier transform of $g(p)$. As a distribution on such functions $g(p)$, the assertion $(19)$ then follows.
\par The structure of the proof outlined above emerges due to the factorization possible in the exponential function. For example, in a simpler case, applying the same method to the integration of an arbitrary well-behaved function on the manifold, not necessarily compact, we could write
$$\int d^4x {\sqrt g}\ f(x) = \Sigma_B \int_Bf(x_B + \eta){\sqrt g(x_B)}\  d^4\eta, \eqno(26)$$
where we again cover the spacetime, assumed geodesically complete, with small boxes related by parallel transport.
\par Since for a small interval $\xi^\lambda$ in $B$,
$$\eta^\mu= {\partial x^\mu \over \partial \xi^\lambda} \xi^\lambda= {a^\mu}_\lambda \xi^\lambda, \eqno(27)$$
 as above, $d^4\eta = {\det a} d^4 \xi$ and we have (in each box $B$, $g= -\det g(B)= (\det a)^{-2}$)
$$ \int d^4x {\sqrt g}\ f(x) =\Sigma_B \int_B f(x_B + {a^\mu}_\lambda\xi^\lambda) d^4\xi. \eqno(28)$$
To lowest order, this is
$$\int d^4x {\sqrt g}\ f(x) =\Sigma_B \int_B f(x_B) d^4\xi, \eqno(29)$$
just our usual understanding of the meaning of $\int d^4x {\sqrt{g}}\ f(x)$ as a sum over the whole space with local measure $d^4 x {\sqrt{g}}$.
\bigskip
\noindent{\bf 2. Consequences for the Quantum Theory}
\bigskip
\par The scalar product for the SHPGR Hilbert space [1] is
$$ \int d^4x \sqrt{g(x)}\  \psi^* (x)\chi(x)= <\psi|\chi>. \eqno(30)$$
As pointed out in [1], the operator $-i {\partial\over \partial x^\mu}$ is not self adjoint in this scalar product. However,  the operator
$$ p_\mu = -i {\partial \over \partial x^\mu} - {i \over 2}{1 \over \sqrt{g(x)}}{\partial \over \partial x^\mu} \sqrt{g(x)}  \eqno(31)$$
is essentially self-adjoint. It was furthermore pointed out that in the  representation obtained by replacing all wave functions $\psi(x)$ by $g(x)^{ 1\over 4}\psi(x)$, which we call the Foldy-Wouthuysen representation in coordinate space[1][15], the operator $(31)$ becomes just $-i {\partial\over \partial x^\mu}$.
\par To cast our results in the familiar form of the quantum theory, we write the scalar product $(30)$ as
$$ <\psi|\chi> = \int d^4 x <\psi|x> <x|\chi>,  \eqno(32)$$
where
$$\eqalign{<x|\chi> &= {g(x)}^{{1\over 4}} \chi(x) \cr
  <x|\psi> &= g(x)^{{1\over 4}} \psi(x), \cr} \eqno(33)$$
(and $<\psi|x> = <x|\psi>^* $) consistently with  $(32)$. This definition coincides with the Foldy-Wouthuysen representation as defined in ref.[1]. We now wish to show that the Parseval-Plancherel relation[9][10] holds for the momentum representation for the integral $(32)$.
\par As in the definition of Fourier transforms given in ref.[1], we define \footnote{*}{We have changed the signs in the exponents appearing in ref.[1] to conform with the usual convention.}
$$ <x|p>  = {1\over (2\pi)^4 {g(x)}^{1\over 4}} e^{ip_\mu x^\mu} \eqno(34)$$
and
$$<p|x> ={g(x)}^{1 \over 4} e^{-ip_\mu x^\mu}, \eqno(35)$$
which also follows from considering the ket $|p>$ as a limiting case of a sharply defined function ${\tilde f}(p)$ in $(6)$ (but in Foldy-Wouthuysen representation). With $(35)$ we have
$$\int d^4p <x|p> <p|x'> = \delta^4 (x-x'). \eqno(36)$$
\par It then follows from $(35)$ that
$$\eqalign{<p|\chi> &= \int d^4 x <p|x><x|\chi> \cr
  &= \int d^4 x\  {g(x)}^{1 \over 4}\  e^{-ip_\mu x^\mu} {g(x)}^{{ 1\over 4}} \chi(x) \cr
  &= \int d^4 x \ e^{-ip_\mu x^\mu}\sqrt{g(x)} \chi(x)= {\tilde \chi} (p).\cr} \eqno(37).$$
  Moreover, from $(34)$, 
$$\eqalign{ <\psi|p> &= \int d^4 x <\psi|x><x|p> \cr
  &= \int d^4 x \ {g(x)}^{{ 1\over 4}}\psi^*(x){1\over 2\pi^4 {g(x)}^{1 \over 4}} e^{ip_\mu x^\mu} \cr
  &= \int {d^4 x\over (2\pi)^4} \ e^{ip_\mu x^\mu}\psi^*(x) \neq {\tilde \psi} ^*(p).\cr}\eqno(38)$$
 Note that this is the complex conjugate of $<p|\psi>$ only in the flat space limit, reflecting the structure of $(34)$ and $(35)$. This function, however, serves as the {\it dual} of the function $<p|\psi>$ for the construction of the scalar product contracting, for example, with $<p|\psi>$ to give the squared norm.
\par  From $(37)$ and $(38)$, we have
$$\eqalign{\int d^4 p <\psi|p><p|\chi> &= \int d^4 p\int {d^4 x\over (2\pi)^4} \cr
&\times e^{ip_\mu x^\mu}\psi^*(x)\int d^4 x'\  e^{-ip_\mu x'^\mu}\sqrt{g(x')} \chi(x')\cr
&= \int d^4x \sqrt{g(x)}\  \psi^* (x)\chi(x). \cr} \eqno(39)$$
 \par This completes our explicit proof of the Parseval relation
$$ \int d^4 x \sqrt{g(x)}\  |\psi(x)|^2 = \int d^4p <\psi|p><p|\psi>.\eqno(40)$$
Note that $<\psi|p><p|\psi>$ is not necessarily a positive number; only the integral assures positivity and unitarity of the Fourier transform, since in this representation, $<\psi|p>$ is not the complex conjugate of $<p|\psi>$.
\par As pointed out above, the operator $p_\mu = -i {\partial\over \partial x^\mu}$
is essentially self-adjoint in the Foldy-Wouthuysen representation. We now examine its spectrum. We use the notation $\{X\}$ and $\{P\}$ to distinguish the canonical operators from the numerical parameters.
\par Since, by definition, we should have
$$ <x|P_\mu |\psi> = -i {\partial \over \partial x^\mu} <x|\psi> \eqno(41)$$
we have, by completeness of the spectral family of $X$,
$$ P_\mu|\psi> = \int d^4x |x>\bigl( -i {\partial \over \partial x^\mu}\bigr) <x|\psi>, \eqno(42)$$
giving $P$ in operator form in the $x-$representation. In $p-$representation, we have\footnote{*}{ In $(43)$, $<x|p'>$ corresponds to a wave function which we might call $\psi_{p'}(x)$ and $<p|x>$ to the dual function ${\psi^\dagger}_p (x)$; in $(45)$, $<p|x'>$ corresponds to what we might call $\psi_{x'} (p)$ and $<x|p>$ to the dual function ${\psi^\dagger}_x(p)$.}   
$$\eqalign{\int d^4x <p|x>P_\lambda <x|p'> &= \int d^4x\  e^{-ip_\mu x^\mu}\bigl({g(x)}^{1\over 4} P_\lambda  {g(x)}^{-{1 \over 4}}\bigr) e^{ip'_\mu x^\mu} \cr
&= p_\lambda \delta^4 (p-p'), \cr} \eqno(43)$$
where we recognize the central factor in parentheses as the Foldy-Wouthuysen form 
of the momentum operator.
\par Finally,in the same way, for the canonical coordinate, we should have
$$ <p|X^\mu|\psi> = i {\partial \over \partial p_\mu} <p|\psi>. \eqno(44)$$
Then, in the $x$ representation ($X^\lambda$ commutes with $g(x)$),
$$ \eqalign{ \int d^4p <x|p> X^\lambda <p|x'> &= \int d^4p{1 \over (2\pi)^4}e^{ip_\mu x^\mu} X^\lambda e^{-ip_\mu x'^\mu}\cr
&= x^\lambda \delta^4 (x-x'). \cr} \eqno(45)$$
\par We now turn to linear superpositions, which have the same form as in the flat space theory. Orthogonal sets (on the measure $d^4x \sqrt{g(x)}$) can be generated using the scalar product $(1)$ for the Schmidt orthogonalization process (or the method of Murray[16] using minimal distance, here defined by choice of $a$ in $\Vert \psi -a\chi \Vert$) to define the orthonormal property
$$ <\phi_n|\phi_m> = \delta_{mn}. \eqno(46)$$
Then, for any linear superposition
$$\psi = \Sigma a_n \phi_n  \eqno(47)$$
we have, as usual,
$$ a_n = <\phi_n|\psi>, \eqno(48)$$
and, if $\Vert \psi\Vert^2 =1, $
$$ \Sigma |a_n|^2 =1,\eqno(49)$$
and $|a_n|^2$ is the probability (Born)to find the system in the state $\phi_n$.
\bigskip
\noindent{\bf 3. Quantum Field Theory}
\bigskip
\par To define a quantum field theory on the curved space, we shall construct a Fock space for the many body theory in terms of the direct product of single particle states in momentum space [4], and define creation and annihilation operators [17]. The Fourier transform of these operators is then used to construct the quantum fields. We have seen in the previous section that for the state $\psi(x)$ of a one particle system, the complex conjugate of the state (we suppress the tilde in following) in momentum space
$$ \psi (p)= \int d^4 x \ e^{-ip_\mu x^\mu}\sqrt{g(x)} \psi(x) \eqno(50)$$
is not equal to the dual $<\psi|p>$ 
$$  \psi^\dagger(p) = {1\over (2\pi)^4} \int d^4x\  e^{ip_\mu x^\mu}\psi^*(x). \eqno(51)$$
Here, the dagger is used to indicate the vector {\it dual} to $\psi(x)$,
necessary, as in Eq. $(39)$, to form the scalar product.  In this form, $(39)$ can be written as
$$\int d^4 p {\psi_1}^\dagger(p) \psi_2 (p)=  \int d^4x \sqrt{g(x)}\  {\psi_1}^* (x)\psi_2(x).  \eqno(52)$$
In this sense, $<p|x>$ (of $(35)$) is the dual of the (generalized) momentum state $<x|p>$ in the Foldy-Wouthuysen representation. The operator representations $(43)$ and $(45)$ are therefore bilinears in the states and their duals, and, as shown below, correspond to the second quantized form of the operators, as in the usual form of ``second quantization''. Note that the linear functional $L(\psi)$ of the Riesz theorem[18] that reaches a maximum for a given $\psi_0$ is given uniquely by the scalar product $(52)$, $L(\psi) = \int d^4p {\psi^\dagger(p)}_0 \psi(p)$.   
\par The many-body Fock space is constructed [17][19] by representing the $N$-body wave function for identical particles on the basis of states of the form, here suitably symmetrized for Bose-Einstein or Fermi-Dirac statistics at equal $\tau$,\footnote{*}{Other approaches to quantum field theory on the manifold of general relativity, such as in [20][21][22][23], introduce a timelike foliation
of space time to describe the fields and their evolution. There is no necessity for us to do this since we have available the universal invariant parameter $\tau$. The usual specification of a spacelike surface on which a complete set of local observables $\{{\cal O}_\tau (x)\} $ commute is then correlated to this $\tau$.} In the folowing we work out the Fermi-Dirac case explicitly;the Bose-Einstein formulation is similar.  We define, for the Fermi-Dirac case, 
$$ \Psi_{N,\tau}(p_N,p_{N-1}, \dots p_1) = {1 \over N!}\Sigma (-1^P P\  \psi_N \otimes \psi_{N-1}\otimes \dots \otimes \psi_1)(p_N,p_{N-1} \dots p_1) , \eqno(53)$$
where all states in the direct product are at equal $\tau$ (with {\it e.g.} $\Psi_2 = {1 \over 2}(\psi_2 \otimes \psi_1- \psi_1\otimes \psi_2)(p_2,p_1) = {1\over 2} (\psi_2(p_2) \psi_1(p_1)-\psi_1(p_2) \psi_2(p_1))$). We work initially in momentum space, since in this representation the structure is most similar in form to the usual construction.  
\par The Fock space consists of span of the set of the form $(53)$, for every $N=(0,1,\dots
\infty)$, where $N=0$ is the vacuum state. We now define the creation operator $a^\dagger(\psi_{N+1}$ on this space with the property that\footnote{*}{Here the dagger indicates the Hermitian conjugate in the Fock space scalar product, as in Eq. $(61)$. We use this notation because, as in $(62)$, its Fock space Hermitian conjugate carries the dual vector to the scalar product $(62)$.}
$$ \Psi_{N+1}(p_{N+1}. p_N, \dots p_1) = a^\dagger(\psi_{N+1}) \Psi_N)(p_N,p_{N+1},\dots p_1)   , \eqno(54)$$
which carries out as well the appropriate antisymmetrization. In order to define the annihilation operator we take the scalar product of this state with an $N+1$ particle
state
$$ \Phi_{N+1,\tau}(p_{N+1},p_N, \dots p_1) = {1 \over (N+1)!}\Sigma (-1^P P \phi_{N+1}\otimes \phi_N\otimes \dots \otimes \psi_1)(p_{N+1},p_N \dots p_1) , \eqno(55)$$
for which
$$(\Phi_{N+1},a^\dagger(\psi_{N+1}) \Psi_N)= (a(\psi_{N+1})\Phi_{N+1}, \Psi_N) \eqno(56)$$
where $a(\psi_{N+1})$, the Hermtian conjugate of $a^\dagger(\psi_{N+1})$ in the Fock space, is an annihilation operator that removes the particle in the state $\psi_{N+1}$.  This scalar product is defined on the momentum space by $(52)$ term by term, using the dual vectors $\psi^\dagger$, as in $(52)$, thus defining the adjoint on the Fock space. For example, for $N=2$,
$$\Psi_2 = {1 \over 2!} (\psi_2 \otimes \psi_1 - \psi_1 \otimes \psi_2). \eqno(57)$$
Then,
$$\eqalign{a^\dagger(\psi_3) \Psi_2 &= {1 \over 3!} (\psi_3 \otimes \psi_2 \otimes \psi_1 +\psi_1 \otimes \psi_3 \otimes \psi_2+\psi_2 
\otimes \psi_3 \otimes \psi_1\cr &-\psi_3 \otimes \psi_1 \otimes \psi_2-\psi_2 \otimes \psi_3 \otimes \psi_1-\psi_1 \otimes \psi_2 \otimes \psi_3).\cr}\eqno(58)$$
We then take the scalar product with
$$\Phi_3 = {1\over 3!} \Sigma_P (-1)^P P \ \phi_3 \otimes \phi_2 \otimes \phi_1, \eqno(59)$$
with conjugate states in the dual space, for which, by the Parseval result,
$$(\phi,\psi)= \int d^4 p \ \phi^\dagger  (p) \psi(p) =  \int d^4x \sqrt{g(x)}\  \phi^*(x)\psi(x). \eqno(60)$$
It then follows, by carrying out the scalar product and selecting terms proportional to the two-body states $\Psi(\phi_i,\phi_j)$, that the action of the operator $a(\psi_3)$ on the state
$\Phi(\phi_3,\phi_2,\phi_1)$ is given by
$$ a(\psi_3)\Phi(\phi_3,\phi_2.\phi_1) = (\psi_3,\phi_3) \Phi_2( \phi_2,\phi_1) -(\psi_3,\phi_2) \Phi_2 (\phi_3,\phi_1) + (\psi_3,\phi_1)\Phi_2(\phi_3,\phi_2), \eqno(61)$$
{\it i.e.}, the annihilation operator acts like a derivation with alternating signs due to its fermionic nature.
\par This calculation has a direct extension to the $N$-body case. For bosons, the procedure may be carried out in a similar way.
\par Applying these operators to  the $N$ and $N+1$ particle states, one finds directly the commutation relations
$$ [a(\psi), a^\dagger (\phi)]_\mp = (\psi,\phi), \eqno(62)$$
so that for orthonormal states (with scalar product $(60)$),
$$[a(\phi_n), a^\dagger (\phi_m)]_\mp= \delta_{nm}. \eqno(63)$$
Based on the Dirac form $(34)$ (in Foldy-Wouthuysen representation), consider the distorted ``plane wave''
$$ {\hat \phi}_p(x) = {1 \over (2\pi)^4 g(x)^{1\over 4} } e^{ip_\mu x^\mu} \eqno(64)$$
and its dual
$$ {{\hat \phi}_p}^\dagger(x) =  g(x)^{1\over 4}  e^{-ip_\mu x^\mu} \eqno(65)$$
so that
$$ ({\hat \phi}_p, {\hat \phi}_{p'}) = \delta^4(p-p') \eqno(66)$$
and
$$[a({\hat \phi}_p), a^\dagger({\hat \phi}_{p'})]_\mp = \delta^4(p-p'). \eqno(67)$$
We may call these operators, as is usually done,  $a(p), a^\dagger (p')$. Then, the Fourier transform with kernel $(34)$ to transform $a(p)$ and $(35)$ to transform the  dual operator $a^\dagger(p)$, we find for the corresponding {\it quantum fields}
$$\eqalign{\psi(x) &=  {1 \over (2\pi)^4 g(x)^{1\over 4} }\int d^4p\  e^{ip_\mu x^\mu}a(p) \cr
\psi^\dagger (x) &=  g(x)^{1\over 4}\int d^4p\   e^{-ip_\mu x^\mu} a^\dagger (p). \cr} \eqno(68)$$
 We have used here the same symbol as for the wave functions and their dual (with the factor $g(x)^{1\over 4}$ for the Foldy-Wouthuysen representation) to maintain a close analogy to the usual form of second quantization (as in $(43)$, $(45)$ and the associated footnote) for the representation of operators on the Fock space. With the commutation relations $(62)$ and the result $(2)$ proven above, it is easy to see that these fields satisfy (at equal $\tau$)
$$ [\psi(x), \psi^\dagger(x')]_\mp = \delta^4(x-x'), \eqno(69)$$
as for the commutation-anti-commutation relations of the usual quantum field theory on Minkowski space.

\bigskip
\noindent{\bf 4. Conclusions}
\bigskip
\par We have constructed a proof of the Fourier transform used in ref.[1], valid for any non-compact geodesically complete manifold. This proof is valid, for example, for the exterior region of the Schwarzschild solution since there is an infinite redshift at the singularity, and in the interior region as well, since the geodesics approach the singularity only asymptotically.  We have, furthermore, extended the discussion of ref.[1] to prove the Parseval-Plancherel theorem, assuring the equality of the norm in both coordinate and momentum representations using the Foldy-Wouthuysen configuration representation discussed there, for convenience, and for the sake of the correspondence of the resulting formulas with those of the usual quantum theory. Following our formulation of the Dirac form of the quantum theory, we identify the functions corresponding to the {\it state} of the system and their dual vectors permitting us to construct the Fock space and the quantum fields for Bose-Einstein and Fermi-Dirac particles on the curved space.
Although the theory is not manifestly diffeomorphism covariant (due to the structure of the Fourier transform), it is indeed invariant in form under arbitrary coordinate transformations as well as in any arbitrary coordinatization of the manifold. 
\par We treat the structure of the theory with representations of particles with spin in a succeeding paper.      

\bigskip
\noindent {\it Acknowledgements}
\par I wish to thank Moshe Chaichian, Asher Yahalom, and Gil Elgressy for fruitful discussions. 
\bigskip
\noindent {\it References}
{\obeylines \smallskip
\frenchspacing

\item{1.} L.P. Horwitz, European Physical Journal Plus, {\bf 134}, 313 (2019).
\item{2.} E.C.G. Stueckelberg, Helv. Phys. Acta {\bf 14}, 372,585 (1941), {\bf 15}, 23 (1942).
\item{3.} L.P. Horwitz and C. Piron, Helv. Phys. Acta {\bf 66}, 316
(1973).
\item{4.} Lawrence Horwitz, {\it Relativistic Quantum  Mechanics},  Fundamental Theories of Physics 180, Springer, Dordrecht (2015).
\item{5.} R.E. Collins and J.R. Fanchi, Nuovo Cim. {\bf 48A},
314 (1978).
\item{6.} J.R. Fanchi, {\it Parametrized Relativistic Quantum
Theory}, Kluwer, Dordrecht (1993).
\item{7.} R. Abraham, J.E. Marsden and T. Ratiu, {\it Manifolds, Tensor Analysis and Applications}, Applied Mathematical Sciences 75, Springer-Veralag. New York (1988).
\item{8.} T.D. Newton and E. Wigner, Rev. Mod. Phys. {\bf 21}, 400 (1949).
\item{9.} M.-A. Parseval des Che\^nes, Memoires pr\'esent\'es \`a l'Institut des Sciences, Lettres et Arts, par divers savants et lus dans les assembl\'ies. Sciences, Math\'ematiques et Physiques,(Savants \'etrangers)vol. 1, 635-688 (1806).
\item {10.}M. Plancherel, Rendiconti del Circolo Mathematico di Palermo {\bf 30}, 298 (1910).
\item{11.} K. Biswas, arXiv 1802.07236 [math. DG](2018),{\it The Fourier Transform on Negatively Curved Harmonic Manifolds}, arXiv 1905.04112 [math. DG](2019) {\it The Fourier Transform on  Harmonic Manifolds of Purely Exponential Volume Growth}. 
\item{12.} M. Reed and B.Simon, {\it Methods of Modern Mathematical Physics}, 1. Functional Analysis, Academic Press, New York (1972).
\item{13.} Y.Strauss,  L.P. Horwitz,J . Levitan and A. Yahalom, Jour. Math. Phys. {\bf 56} 072701 (2015).
\item{14.} V. Georgiev, {\it Semilinear Hyperbolic Equations}, Tokyo Mathematical Society.  (Chap. 8, Fourier Transform on Manifolds with Constant Negative Curvature), p.126.Japan (2005)
\item{15.} L.L. Foldy and S.A. Wouthuysen, Phys. Rev. {\bf 78}, 29 (1950).
\item{16.}F.J. Murray. {\it Linear Transformations in Hilbert Space}, Princeton University Press, Princeton (1941).
\item{17.} G. Baym, {\it Lectures on Quantum Mechanics}, W.A. Benjamin, New York (1969).
\item{18.} F. Riesz and B.Sz. Nagy,{\it Functional Analysis}, Frederick Ungar, New York (1955).
\item{19.} L.P. Horwitz and R.I. Arshansky, {\it Relativistic Many-Body Theory and Statistical Mechanics}, IOP, Bristol; Morgan \& Claypool, San Rafael (2018).
\item{20.}.N.D. Birrell and P.C.W. Davies,{\it Quantum Fields in Curved Space},  Cambridge University Press,Cambridge (1982)
\item{21.} E. Poisson, Cambridge University Press, Cambridge (2004).
\item{22.} B.S. DeWitt, Rev.Mod.Phys.{\bf 29},377 (1957)
\item{23.} A.V. Stoyanovski, arXiv:0910.2296 (2009).

\end